\documentclass[11pt, titlepage]{article}
 \usepackage[super]{natbib}
 \usepackage{amsmath} 
 \usepackage{enumerate}

 \renewcommand{\section}[1]{\medskip \addtocounter{section}{1}
    \noindent\textbf{\Roman{section}. \ #1}\medskip \setcounter{subsection}{0}
     \parindent=5ex}
 \renewcommand{\subsection}[1]{\medskip \addtocounter{subsection}{1}
    \textbf{\Alph{subsection}. \ #1} \medskip \setcounter{subsubsection}{0}}

\begin{document}

 \begin{titlepage}

 \begin{center}

 \textbf{Derivation of Maxwell's equations via the}\\
 
 \textbf{covariance requirements of the special theory of relativity,} \\
 
 \textbf{starting with Newton's laws}\\

 \vspace{10ex}

 Allan D. Pierce\footnote{e-mail: adp@bu.edu}\\

 Boston University \\

 Boston, Massachusetts

 \end{center}

 \end{titlepage}

\begin{abstract}

 A connection between Maxwell's
 equations, Newton's laws, and the special theory of relativity is
 established  with a derivation that begins with
 Newton's verbal enunciation of his first two laws.  Derived equations 
 are required to be covariant, and  
a simplicity criterion requires that the four-vector force on a charged particle
 be linearly related to the four-vector velocity.  The connecting 
 tensor has derivable symmetry properties and contains the
 electric and magnetic field vectors.   The Lorentz force law 
 emerges, and  Maxwell's equations for free space
 emerge with the assumption that the tensor and its dual must
 both satisfy first order partial differential equations.  The inhomogeneous  extension yields a charge
 density and a current density as being the source of the field,
 and yields the law of conservation of charge.    Newton's third law is
 reinterpreted  as a reciprocity statement, which requires that the charge in the source term can be taken as the same physical entity as that of the test particle and that both can be assigned the same units.   Requiring covariance under  either spatial inversions or time reversals precludes  magnetic charge being a source of electromagnetic fields that exert forces on electric charges. 
  
\end{abstract}

\section{Introduction}

\label{intro}
Maxwell's equations and the special theory of relativity are
 intimately related, and a rich literature exists that explores and
 elucidates this connection.  One major theme is that one can
 start with the basic ideas of the theory of special relativity and,
 with some basic experimental laws and with a small number of
 intuitively simple assumptions,  derive Maxwell's equations.
 Another theme is that one can start with two or more relatively
 simple physical  laws, not explicitly relying on  special relativity, and 
 once again derive Maxwell's equations.  
 
 The principal theme of the present paper is that there is a strong
 connection between Newton's laws and Maxwell's equations, and that
 this connection is provided by the special theory of relativity.  
 The viewpoint here is wholly classical,  although not  mechanistic.
 The electrical and magnetic fields are not regarded as mechanical
 entities, but as classical fields which are created by moving charges and which exert forces on moving charges.  A suitable reinterpretation  of Newton's laws is assumed to apply to the masses  that 
 experience forces caused by the fields, and the fields are  assumed
 to be governed  by an independent set of equations.   Use is made
 of Einstein's two postulates\cite{Einstein1905}${}^,$\cite{DoverRelativity}${}^,$\cite{Kilmister} that  (i) the  equations
 must have the  same form  in equivalent coordinate systems  and that
(ii)   the speed of light must be  the  same in all such coordinate systems.
 The concept of covariance is applied in the sense of Minkowski\cite{Minkowski1908},
  so that equivalent coordinate systems are taken to be those where the coordinates
 of one are related to those of another by  a Lorentz transformation.
   
  In regard to Newton's laws being used as a starting point, the idea of such goes at least as far back as 1948, when Feynman\cite{Dyson} showed Dyson a ``proof'' of Maxwell's equations ``assuming only Newton's laws of motion and the commutation relation between position and velocity for a single nonrelativistic particle.''  That proof as reported by Dyson is  perplexing, as it is difficult to see how a set of equations that predict propagation at a speed with a precise unique value should result from a formulation that does not explicitly involve the speed $c$ of light.  Feynman was using units in which $c$ was numerically equal to unity, and the details of his thinking are encoded in the remark that the ``other two Maxwell equations merely define the external charge and current densities.''  The treatment in the present paper follows more traditional lines of thinking and is limited entirely to the realm of classical physics, given the normally accepted inclusion of the special theory of relativity in classical physics.  Nevertheless, as is discussed further below, Feynman's provocative remark supplies a crucial hint as to what should be an appropriate reinterpretation of Newton's third law.
 
 In regard to the more traditional treatments appearing in previous literature, one should first note that 
Maxwell's equations preceded the special theory of relativity in the history of physics, and one might loosely state that relativity developed because of the need to insure that Maxwell's equations be independent of any relative velocity between coordinate systems.  But, 
after the emergence of relativity as a fundamental cornerstone of physics, papers and books began to appear that ``derived'' Maxwell's
equations.   A major category of such treatments takes Coulomb's
law, or equivalently, the ``laws of electrostatics'' as a starting point.
The earliest such derivation was given by Page\cite{Page1912} in
1912, and that treatment was subsequently refined in the 1940
textbook by Page and Adams\cite{PageandAdams}. 
Frisch and Wilets\cite{Frisch} in a 1956 paper  criticize Page and Adams, stating that they use ``an 
apparently overspecialized model: an emission theory of lines of force,'' and give an alternate derivation, making a series of plausible
(but not manifestly obvious) postulates, which can be construed as including Coulomb's law.  They also bring attention to a 1926 paper by
Swann\cite{Swann} where two derivations, involving the use of invariance under Lorentz transformations,  of equations resembling Maxwell's equations are given.  
 
 More recent treatments making use of Coulomb's law were given by Elliott\cite{ElliottIEEE} and Tessman\cite{Tessman} in 1966, and a brief pedagogical development was given by Krefetz\cite{Krefetz} in 1970, who drew attention to  Feynman's remark\cite{Feynmanbook}, ``it is sometimes said, by people who are careless, that all of electrodynamics can be deduced solely from the Lorentz transformation and Coulomb's law,'' which is followed by statements to the effect that it is always necessary to make some additional assumptions.  Krefetz pointed out that ``what constitutes a reasonable assumption is, after all, a matter of taste.''

Another theme for the derivation of Maxwell's equations can be traced back to Landau in 1933.  Podolsky, in the preface of his text with
Kunz\cite{Podolsky}, refers to discussions he had
with L. D. Landau in 1933 on the goal of ``presenting classical
electrodynamics as theory based on definite postulates of a
general nature, such as the principle of superposition, rather
than [inductively inferring  the theory from] experimental laws.''  Thus, in the venerable Landau and Lifshitz series\cite{LandauFields}, one finds an elegant and extensive development, which begins with the assumption of the existence of a four-potential, which presumes the validity of the principle of least action (Hamilton's principle) in which the time and spatially varying potentials are treated as generalized coordinates, and which makes a series of plausible assumptions concerning the form of the action function.   The treatment is an intricate blend of sophisticated mathematical constructions of theoretical physics and plausible assumptions, although in a footnote the authors state: ``The assertions which follow should be regarded as being, to a certain extent, the consequence of experimental data.  The form of the action for a particle in an electromagnetic field cannot be fixed on the basis of general considerations alone.'' 

It would unduly lengthen the present paper if one attempted to discuss, even in a cursory manner, all the papers and book passages that have been concerned with the derivation of Maxwell's equations,
and the present author cannot claim to have seen all those that are currently available, let alone digested them.  Among those that should be mentioned are a sequence of papers by Kobe\cite{Kobe1978}${}^,$\cite{Kobe1980}${}^,$\cite{Kobe1984}${}^,$\cite{Kobe1986} which examine the topic from a variety of perspectives and which also give extensive references.  Other papers of interest are those by 
Crater\cite{Crater}, Jefimenko\cite{Jefimenko1996}, Ton\cite{Ton},
Griffiths and Heald\cite{Griffiths}, Crawford\cite{Crawford},
Neuenschwander and Turner\cite{Neuenschwander},  Bork\cite{Bork},Goedecke\cite{Goedecke},
and Hokkyo\cite{Hokkyo}.

The manner in which the present paper's development differs from what has appeared previously in the literature is addressed more fully further below and in the concluding remarks section.

\section{Relativistic version of Newton's second law} 
\label{Rel_version}

\setlength{\parskip}{5pt}

 The discussion here begins, interlacing a brief summary of some basic tenets of the special theory of relativity, with a concise derivation of the
covariant form, first given by Minkowski\cite{Minkowski1908}, of
Newton's second law.  The derivation differs from what has previously been published in that it specifically draws on Newton's verbal enunciation\cite{Newton} of his first two laws.
    
\par    One seeks a description for the evolution
of the space-time coordinates of a test particle in a ``Minkowski''
space\cite{Minkowski1909}${}^,$\cite{DoverRelativity}${}^,$\cite{Bergmann} with (\emph{world point}) coordinates $X^1 = x$, $X^2 = y$, $X^3 = z$, and $X^4 =ct$, 
 where $c$ is 
the speed of light.  The coordinates of the particle itself are
distinguished by a subscript $P$ (for particle).
 Whatever equations are derived are required to be the same
 in any one of an equivalent set of coordinate systems, these
 being such that the speed of light is the same in each such
 system. 
 
 Suppose, for example, that  $\Delta X^\alpha$
 is a set of coordinate  increments in one admissible coordinate system,
 with  the spatial separation equal to $c$ times the time separation,
so that
 \begin{equation} \label{firstincrement}
- (\Delta X^1)^2 - (\Delta X^2)^2 - (\Delta X^3)^2 + (\Delta X^4)^2 = \Delta X^\alpha g_{\alpha\beta}
\Delta X^\beta =  0.
\end{equation} 
(The second version here makes use of common tensor 
notation\cite{EinsteinGrundlage1916}${}^,$\cite{Bergmann}, with $g_{\alpha\beta}$
being the metric tensor, a diagonal matrix with diagonal elements $-1$, $-1$, $-1$,
and $+1$.)    Then an analogous relation must hold for a second coordinate system,
so that
 \begin{equation} \label{secondincrement}
- (\Delta Y^1)^2 - (\Delta Y^2)^2 - (\Delta Y^3)^2 + (\Delta Y^4)^2 = \Delta Y^\alpha g_{\alpha\beta}
\Delta Y^\beta =  0.
\end{equation} 
Admissible transformations that connect two such coordinate
systems are taken to  be
linear relations, so that one can write, for an arbitrary set of increments (T for transformed),
\begin{equation}\label{Lorentztransf}
\Delta X_T^\alpha =\Delta Y^\alpha = \Lambda^\alpha_{{}\beta} \Delta X^\beta,
\end{equation}
where the transformation matrix $\Lambda^\alpha_{{}\beta}$ is independent of the
coordinates.  Given this relation, a brief derivation shows 
that Eq.~(\ref{secondincrement}) follows from Eq.~(\ref{firstincrement}) provided
the transformation matrix satisfies the relation\cite{Bergmann}${}^,$\cite{Low}
\begin{equation}\label{defininga}
 \Lambda^\gamma_\alpha
g_{\gamma\delta}\Lambda^\delta_\beta
=  g_{\alpha\beta}. 
\end{equation}
There is a wider\cite{Cunningham}${}^,$\cite{Bateman} class of transformations that leaves the speed of light unchanged, 
but the class represented by the above provides sufficient guidance for identification
of a covariant theory.  The transformations allowed by this relation include 
rigid body rotations, time reversals, spatial inversions, Lorentz's
and Einstein's transformation between moving coordinate systems, and any
arbitrary sequence of these.   Following Poincare\cite{Poincare1906}${}^,$\cite{Kilmister}${}^,$\cite{SchwartzRendiconti},  such transformations are
here referred to as Lorentz transformations, and they form a group.  The determinant of any matrix satisfying Eq. (\ref{defininga}) can be either $+1$
or $-1$, and one can also show\cite{Weinberg} that if $\Lambda^4_4>0$ for each of two consecutive transformations, then this is so for the combined transformation.  The subgroup for which the determinant is $+1$ and for which $\Lambda^4_4>0$ is known as the \emph{proper orthochronous Lorentz
subgroup}.  The full group includes this subgroup plus the time reversal
transformation (diagonal with elements 1, 1, 1,  and $-1$) and the spatial
inversion transformation (diagonal with elements $-1$,$ -1$,$-1$,
and $+1$) and all of the products formed from
the subgroup and these two operators.  One can also conceive of subgroups
which exclude the time-reversal operator and which exclude the spatial-inversion operator.  This feature of the full Lorentz group leaves open the question of whether the equations of classical physics should be invariant under time reversals or
under spatial  inversions.   The mathematical structure allows, in regard to the inclusion of these two operators,  the following choices:  (i) neither, 
(ii) only one of the two,  (iii) both, but only if simultaneous, and
(iv) both, regardless of whether simultaneous or individual.   

The defining property, Eq. (\ref{defininga}), of the Lorentz transformation allows a definition\cite{Minkowski1909} of a proper time $\tau_P$ for a point particle that is an invariant with respect to the orthochronous proper subgroup and with respect to spatial inversions, but which changes sign under time reversals.
The trajectory of the test particle is described in parametric form with
 the space-time coordinates $X_P^\alpha$  all regarded to be functions
 of a parameter $\tau_P$, defined so that time is a monotonically increasing
 function of $\tau_P$, and so that increments of $\tau_P$ can be computed from
 \begin{equation}\label{deetausquared}
  c^2 (d\tau_P)^2 =  c^2 (dt_P)^2 -(dx_P)^2 - (dy_P)^2 - (dz_P)^2 = 
 d X_P^\alpha g_{\alpha\beta} dX_ P^\beta.
 \end{equation}
 The form of the second expression justifies the assertion that $(d\tau_P)^2$ is fully invariant.  Equivalently, since $d\mathbf{x}_P = \mathbf{v}_P dt_P$, one can
 express the relation between $t_P$ and $\tau_P$ in any given coordinate
 system as 
 \begin{equation}
 d\tau_P = (1/\kappa_P) dt_P; \qquad \kappa_P = [1-(v_P/c)^2]^{-1/2}.
 \end{equation}
 where $\mathbf{v}_P$ is the particle's velocity.
 
 The chief feature of this proper time is that it allows one to identify a
 tentatively suitable four-vector counterpart of the particle velocity as
 $U_P^\alpha = dX_P^\alpha/d \tau_P$.  Given the stated transformation properties of the differential $d\tau_P$, this four-vector
 transforms under the full Lorentz group as
  \begin{equation} \label{Lorentzwithsign}
  U^\alpha_{P,T} = \textrm{sign}\left(\Lambda^4_4\right)\Lambda^\alpha_\beta
  U_P^\beta.
  \end{equation}
  Because of the sign-factor, one would say that $U_P^\alpha$ is not a genuine four-vector, but some sort of pseudo-four-vector.  
  (It is referred to in what follows as a \emph{pseudo-four-vector of the
  time-reversal kind}.)  Nevertheless,
  as long as one knows its transformation rule and formulates equations consistent with this, it can be used in a covariant formulation.  In particular,
  one may note that its derivative with respect to proper time is a genuine
   four-vector.
   
 The foregoing
 provides sufficient mathematical structure for the covariant interpretation of
 Newton's first two laws.  These, as originally enunciated (after translation to
 English) by Newton\cite{Newton}, are as follows: 
 \begin{description}

\item{ \textbf{Law 1.} }{\emph{ Every body perseveres in its state of
being at rest or of moving uniformly straight forward, except
insofar as it is compelled to change its state by forces
impressed.}}

\item{\textbf{Law 2.}}  {\emph{A change in motion is  proportional to
the motive force impressed and takes place along the straight
line in which that force is impressed.}}

\end{description}

For the covariant interpretation of these, one makes use of
what is available that transforms properly under Lorentz
transformations.
The covariant statement of the first law is that $U_P^\alpha$
must be a constant if there are no forces.  The covariant
expression for ``change in motion'' is $dU_P^\alpha/d\tau$, and
 the proportionality constant must be what is ordinarily
termed  the ``rest mass'' ($m_o$).  The phrase ``along a straight
line in which the force is impressed'' has to be loosely
interpreted as saying that there is a contravariant vector
termed ``force'' which is ``parallel'' to the ``change in motion''
four-vector.  Thus one arrives at the relation
\begin{equation} \label{Newton4}
 m_o\frac{ dU_P^\alpha}{d\tau_P} = R_P^\alpha, 
\end{equation}
where the right side, the four-vector force (Minkowski force),  is a contravariant vector with
as-yet-undefined components $R_P^\alpha$.  It should be a genuine
four-vector and transform under the full Lorentz group as in 
Eq. (\ref{Lorentztransf}).

\section{The Lorentz force law}
\label{Lorentz}

The four-vector force of interest here is that which is
associated with an electromagnetic field.  A primary
assumption is that there is an ``external'' part of this field
which exists independently of the presence of the test
particle.  Whatever characterizes this field depends only on the
four space-time (Minkowski) coordinates and is independent
of the particle's mass and velocity.  However, the force
exerted by this field may   well depend on the particle's
velocity. Also, one assumes that the particle has an additional
scalar property, a charge $q_P$, which is defined so
that  all the force components exerted by the electromagnetic
field on the particle are directly proportional to $q_P$.  For this
particle, which has no other intrinsic structure, the four-vector
velocity (Minkowski velocity)
$U^\alpha_P$ is the only simple tensor that one has
available for  the formulation of a covariant expression for the
four-vector force.   One can  argue that there is some weak
limit, which probably has very wide applicability, where the
four-vector force is linearly related to the four-vector velocity.
Thus one is led to the plausible postulate (sort of a
Hooke's law of electromagnetism) that
\begin{equation} \label{Hooke}
R_P^\alpha = q_P\Phi^\alpha_\beta U_P^\beta,  
\end{equation}
where covariance requires the entity $\Phi^\alpha_\beta$  be a
tensor-like quantity (one contravariant index and one covariant index) that
transforms appropriately under all Lorentz transformations. 
(The mathematical apparatus of tensor analysis is used here,
with superscripted indices referred to as contravariant indices
and subscripted indices referred to as covariant indices,
and with the metric tensor $g_{\alpha\beta}$ available for
the lowering of indices.)

The actual transformation rule, for the purely contravariant form,
as deduced from Eqs. (\ref{Lorentztransf}) and (\ref{Lorentzwithsign}),
is
\begin{equation}\label{PhiTransform}
\Phi^{\alpha\beta}_T = \mathrm{sign}\left(\Lambda_4^4\right)
\Lambda^\alpha_\mu \Lambda^\beta_\nu \Phi^{\mu\nu},
\end{equation}
where the sign-factor is $-1$ for transformations that involve
time-reversals.  Thus, $\Phi^\alpha_\beta$ is, strictly speaking, not
a  tensor with regard to the full Lorentz group, but some
sort of pseudo-tensor.   
To distinguish this type of pseudo-tensor from other types that
appear further below, it is referred to as a \emph{pseudo-tensor of 
the time-reversal kind}.

The identification in Eq. (\ref{Hooke}) is not unique, and there are
many possibilities, such as a quadratic expression of the form
$q_P \Phi^\alpha_{\beta\gamma} U_P^\beta U_P^\gamma$,
where $\Phi^\alpha_{\beta\gamma} $ is some as yet undetermined
tensor with three, rather than two, indices.  But the expression in
Eq. (\ref{Hooke}) is the simplest of all such expressions.  

Equation (\ref{Hooke}) is of course well-known, but existing
discussions in the literature usually arrive  at it \emph{after}  the
electromagnetic field tensor  $\Phi^\alpha_\beta$ has been 
previously arrived at by other means.  Low\cite{Low}, for example, infers
the electromagnetic field tensor first, with reference to experimental
laws, and then argues that the four-vector force must be linear
in  the electromagnetic field tensor, and then argues that the
only plausible covariant expression has to be of the
 form of Eq. (\ref{Hooke}).
In retrospect, this is very satisfying to one's intuition, but the argument
is not available in the present context, as one is  assuming that
one knows nothing about the electromagnetic field tensor at
this point, other than that it is a tensor that adheres to a definite transformation law.  (In what follows, the term ``tensor'' is used
loosely to refer to both pseudo-tensors and genuine tensors.)

Since the tensor $\Phi^\alpha_\beta$ exists independently 
 of the presence of the charge, it is a continuum field.
  Each component is independent of any parameters
characterizing the particle, but each depends on the space-time
coordinates. The presence of other bodies that affect
the motion of the test particle is presumed to be fully
accounted for by the properties of the field; and such other
bodies are regarded as sources of the field.  (Note that the
four-vector $U_P^\beta$ can never be identically zero, as
there is always a fourth non-zero component.  
There is no contradiction here with the expectation
that a body at rest can experience an electromagnetic force.)
 
The remaining arbitrariness in the tensor $\Phi^\alpha_\beta$
is drastically reduced by a derivable orthogonality condition 
between the four-vector velocity and the four-vector force,
\begin{equation}\label{orthoga}
R_P^\alpha U_{P,\alpha} = R_P^\alpha g_{\alpha\beta} U_P^\beta = 0,
\end{equation}
which was first noticed by Minkowski\cite{Minkowski1908}.  To derive this, 
one multiplies Eq.~(\ref{Hooke}) by $U_{P,\alpha}$, performs the 
implied sum, and recognizes that $U_{P,\alpha} dU_P^\alpha/d\tau_P$
is $(1/2)d(U_P^\alpha U_{P,\alpha})/d\tau_P$.  But the sum
$ U_P^\alpha U_{P,\alpha}$ is a scalar, a constant equal to 
$c^2$, so its derivative is zero, and  Eq.~(\ref{orthoga}) results.

The implication of the orthogonality relation in regard to
Eq.~(\ref{Hooke}) is that
\begin{equation}\label{squeezed}
  U_{P,\alpha}
\Phi^{\alpha\beta} U_{P,\beta}= 0 ,
\end{equation}
 where $\Phi^{\alpha\beta} = \Phi^\alpha_\gamma g^{\gamma\beta}$
is the purely contravariant form (two contravariant indices) of
the mixed tensor $\Phi^\alpha_\beta$.   Although the magnitudes
of the components of $U_{P,\alpha}$ are constrained so that the
inner product of its contravariant and covariant forms is $c^2$,
they are otherwise arbitrary, and the above equation must hold
for all such vectors.  The admissible arbitrariness leads to the
deduction that the purely contravariant form of the electromagnetic
tensor is antisymmetric, so that 
  \begin{equation} \label{antisymmetric}
 \Phi^{\alpha \beta} = - \Phi^{\beta \alpha}.  
\end{equation}

[The proof just given can be discerned, although in a somewhat different context, in the
text by Melvin Schwartz\cite{Schwartz} and in a 1986 paper by Kobe\cite{Kobe1986}.
The relevant passage in Schwartz's text is in a footnote, with the development appearing
there attributed to D. Dorfan.  Kobe gives an explicit derivation of  Eq. (\ref{antisymmetric})
from Eq. (\ref{squeezed}).  The derivation is also given in the paper by 
Neuenschwander and Turner\cite{Neuenschwander}.]

  With the aid of some hindsight regarding the
symbols that one uses to  label the off diagonal elements, the
(antisymmetric) matrix representation (second index corresponding to
columns) can be written with all generality as
\begin{equation} \label{Phimatrix}
[\Phi^{\alpha\beta}] =
\left( \begin{array}{cccc} 0& -B_z & B_y & E_x/c \\ B_z & 0 & -B_x &
E_y/c \\ -B_y & B_x & 0 & E_z/c
\\ -E_x/c & -E_y/c & -E_z/c & 0  \end{array}\right) .
\end{equation}
 Then, since
$\Phi^\alpha_\beta = \Phi^{\alpha\gamma}
g_{\gamma\beta}$, the postulated force relation of Eq.~(\ref{Hooke})
becomes
\begin{equation}  \label{matrixform}
\left( \begin{array}{c}f_{P,x} \\ f_{P,y} \\ f_{P,z} \\ 
\mathbf{v}_P\cdot\mathbf{f_P}/c \end{array}\right)
 = q_P
\left( \begin{array} {cccc} 0& B_z & -B_y & E_x/c \\ -B_z & 0 & B_x &
E_y/c \\ B_y  & -B_x & 0 &E_z/c  \\  E_x/c & E_y/c &
E_z/c & 0 \end{array} \right)
\left( \begin{array}{c} v_{P,x} \\ v_{P,y} \\ v_{P,z} \\ c
\end{array} \right).
\end{equation}
Here $\mathbf{f}_P$ is the three-vector force which appears when 
the first three components of Eq.~(\ref{Hooke}) are written out explicitly
in vector notation as 
 \begin{equation} \label{Newton}
 \frac{d\mathbf{p}_P}{dt} = \mathbf{f}_P, 
 \end{equation}
with the momentum defined as
\begin{equation}
 \mathbf{p}_P = \frac{m_o}{[1 - (v_P^2/c^2)]^{1/2}}\mathbf{v}_P =
m^*_P \mathbf{v}_P .
\end{equation}
(Here $m^*$ is the relativistic mass.)
Also, in Eq.~(\ref{matrixform}), one has identified $R^1_P = \kappa_P f_{P,x}$ with analogous
relations for the 2-th and 3-rd components.  The fourth component,
$R^4_P$, is derived from the orthogonality relation of Eq.~(\ref{orthoga}).

The first three components of Eq.~(\ref{matrixform}) yield, with vector
notation,
\begin{equation} \label{Lorentzforce}
\mathbf{f}_P = q_P(\mathbf{E} + \mathbf{v}_P\times \mathbf{B} ),
\end{equation}
which is the Lorentz force law\cite{Heaviside}${}^,$\cite{Lorentzbook}.  The fourth
component equation is then only a  redundant corollary of the
first three, because
$ \mathbf{v}_P\cdot \left(\mathbf{v}_P\times \mathbf{B} \right) = 0.   $.
 Although one might not expect anything otherwise, it may be
surprising to some that the Lorentz force equation is a direct
consequence of the orthogonality of the four-vector velocity $U_P^\alpha$
and the four-vector force $R_P^\alpha$.  [A
derivation of the Lorentz force equation via the special theory of relativity was apparently first given by Tolman,~\cite{Tolman} where he used the previously derived Lorentz transformation laws of the electromagnetic field components
to infer the force law in a system where the charge was moving with
speed $\mathbf{v}$ from the force law in a system where the charge was momentarily stationary.  In retrospect, the derivation here is equivalent to
that of Tolman, only the requirement of covariance enables one to bypass using the explicit form of any Lorentz transformation.]

With reference to the transformation rules in Eqs. (\ref{Lorentzwithsign})
and (\ref{PhiTransform}) , one deduces that, under pure time-reversals,
\begin{equation}
\mathbf{v}_P \rightarrow -\mathbf{v}_P; \quad \mathbf{E} \rightarrow \mathbf{E} ;\quad 
 \mathbf{B} \rightarrow -\mathbf{B} ; \quad \mathbf{f}_P \rightarrow \mathbf{f}_P \qquad \mbox{(time-reversals)} ,
 \end{equation}
 while under pure spatial-inversions
 \begin{equation}
\mathbf{v}_P \rightarrow -\mathbf{v}_P; \quad \mathbf{E} \rightarrow -\mathbf{E} ; \quad
 \mathbf{B} \rightarrow \mathbf{B} ; \quad \mathbf{f}_P \rightarrow -\mathbf{f}_P \qquad \mbox{(spatial-inversions)} ,
 \end{equation}
 so, given these rules, the Lorentz force relation is fully covariant under time-reversals and spatial-inversions.

[The symbols assigned to the matrix
elements in Eq.~(\ref{Phimatrix}) are appropriate for SI (rationalized
MKS) units, where the components of the vector $\mathbf{E}$ have
the units of volts per meter, or newtons per coulomb, and
where the components of
$\mathbf{B}$ have the units of teslas, or webers per square meter,
or newton-seconds per coulomb-meter.   In the commonly
used Gaussian system of units, distances have units of
centimeters, forces have units of dynes, and charge has the
units of statcoulombs.  The electric field is denoted by
$\mathbf{E}$ and has units of dynes per statcoulomb, and the
magnetic field $\mathbf{B}$ has the units of gauss.  The unit
of charge, the statcoulomb, is defined so that the gauss and
the dynes per statcoulomb have the same units, and this
requires $\mathbf{B}/c$ to replace the SI magnetic field
$\mathbf{B}$ in the Lorentz force equation, so that for Gaussian units Eq.~(\ref{Lorentzforce}) is
replaced by
\begin{equation}
 \mathbf{f}_P = q_P\left( \mathbf{E} + \frac{ \mathbf{v}_P}{c}\times 
\mathbf{B}\right). 
\end{equation}
To render Eqs.~(\ref{Phimatrix}) appropriate for Gaussian units, one need
only replace the quantities $B_x$, $B_y$, and $B_z$, wherever
they appear, by $B_x/c$, $B_y/c$, and $B_z/c$.]

\section{Dual nature of electric and magnetic fields}
\label{Dual_nature}

At this point, it is appropriate to interject a remark
  from an autobiographical essay\cite{Schilpp} written
by Einstein relatively late in his life,

\begin{quote}

\emph{[Maxwell's equations] can be grasped formally in satisfactory fashion only by way of the special theory of relatively.  [They] are the simplest Lorentz-invariant field equations which can be postulated for an anti-symmetric tensor derived from a vector field.}

\end{quote}

\noindent
While Einstein in this essay does not specify  to just which antisymmetric tensor he is referring, one can infer from the development\cite{EinsteinGrundlage1916}${}^,$\cite{DoverRelativity} in his 1916 \emph{Grundlagen} paper [Section 20, Eqs. (59), (60),
and (61)] that the tensor is essentially the same as the
tensor $\Phi^{\alpha\beta}$ that 
appears here in the four-vector force equation, Eq. (\ref{Hooke}).
[The actual relation, when Gaussian units are used, is
that Einstein's $F_{\alpha\beta}$  is $-c\Phi_{\alpha\beta}$,
where the indicated tensors are the purely covariant forms.]
The vector field to which Einstein refers is the
  four-vector potential.  
In what follows, this same tensor is brought into play, and the development manages to 
sidestep any assumption that it has to be derived from a four-vector field.  

The premise here is that
 the tensor $\Phi^{\alpha\beta} $ identified in the previous
section is a natural building block for a covariant formulation
of a set of partial differential equations (Einstein's simplest
Lorentz-invariant field equations) that govern the time
and spatial evolution of the individual elements of the tensor.   
This tensor has six nonzero elements that are possibly
different from each other,  so the description of its evolution
requires at least six equations. One might anticipate at first
that this tensor can yield at most only four equations, which
would be insufficient for a complete formulation.
 However, the antisymmetry of $\Phi^{\alpha\beta} $ and the
specific property in Eq.~(\ref{defininga}) of the Lorentz transformation
allows one to identify\cite{LandauFields} a   ``dual tensor'' as
\begin{equation} \label{dual}
\Psi^{\alpha\beta} = {\scriptstyle\frac{ 1}{ 
2}}d^{\alpha\beta\mu\nu}
\Phi_{\mu\nu},
\end{equation}
which can be used for the derivation of additional equations.
Here the symbol 
$d^{\alpha\beta\mu\nu}$ is defined to be zero if any two of
its indices are numerically equal, and to be unity ($+1$) if the
ordered set of numbers $\alpha$, $\beta$, $\mu$, $\nu$ is an
even permutation of the integers 1, 2, 3, 4.  If  the permutation
is odd, then the value is $-1$.    [This symbol is occasionally, with 
various notations,
referred to as the Levi-Civita symbol\cite{Moller}${}^,$\cite{Levi} and
also as a permutation symbol\cite{Synge}.]

 This definition in Eq.~(\ref{dual}), in conjunction with Eq.~(\ref{Phimatrix}), leads to
an entity which has the matrix representation (columns
labeled by second index)
\begin{equation} 
[\Psi^{\alpha\beta}] =\left( \begin{array}{cccc}0& -E_z/c & E_y/c & -B_x
\\ E_z/c & 0 & -E_x/c & -B_y 
\\ - E_y/c & E_x/c & 0 & -B_z
\\ B_x&  B_y &  B_z & 0 \end{array} \right) . 
\end{equation}
The relationship of Eq.~(\ref{dual}) is accordingly equivalent to the substitutions
\begin{equation}
 \mathbf{B} \rightarrow \mathbf{E}/c; \qquad \mathbf{E}/c \rightarrow -
\mathbf{B}. 
\end{equation}
Alternately, one can produce
$[\Phi^{\alpha\beta}]$ from $[\Psi^{\alpha\beta}]$ with the
reverse of these substitutions. 

To determine the tensorial nature of the entity
$\Psi^{\alpha\beta}$, one first notes that a standard method\cite{Schreier} for calculation of a
determinant yields
\begin{equation}
 \mathrm{det}[\Lambda] =
d^{\alpha\beta\mu\nu}\Lambda^1_\alpha\Lambda^2_\beta
\Lambda^3_\mu\Lambda^4_\nu, 
\end{equation}
where
$\Lambda^\beta_\alpha$ is a given matrix's element in the
$\beta$-th row and $\alpha$-th column.    Moreover, since the sign of a determinant
changes when any two rows are interchanged or when any two
columns are interchanged, and since it is zero when any two
are the same, one has
\begin{equation} \label{contra4}
d^{\alpha\beta\mu\nu}\Lambda^{\alpha'}_\alpha\Lambda^{\beta'}_\beta
\Lambda^{\mu'}_\mu\Lambda^{\nu'}_\nu =
d^{\alpha'\beta'\mu'\nu'} \mathrm{det}[\Lambda]. 
\end{equation}
Because this applies for all matrices, it applies to any matrix
that corresponds to a Lorentz transformation.   Also, because
one wishes the definition of the dual in Eq. (\ref{dual}) to be 
applicable in all equivalent coordinate systems, the entity
 $d^{\alpha\beta\mu\nu}$, if regarded as something analogous to a tensor, is required to be the same in all coordinate systems,
 so that, with a properly devised transformation rule,
 $d_T^{\alpha\beta\mu\nu} =d^{\alpha\beta\mu\nu}$.  While this might in itself be taken as the transformation rule, it is helpful in the derivation of transformation rules for cases such as that of Eq. (\ref{dual})  to express this in a manner analogous to that for a tensor with four contravariant indices.
Such a transformation rule, as deduced from Eq. (\ref{contra4}),
is \begin{equation} \label{ruleforLevi}
d_T^{\alpha'\beta'\mu'\nu'}
=(\mathrm{det}[\Lambda])^{-1} \Lambda^{\alpha'}_\alpha\Lambda^{\beta'}_\beta
\Lambda^{\mu'}_\mu\Lambda^{\nu'}_\nu d^{\alpha\beta\mu\nu}.
\end{equation}
For the special case when the determinant is $+1$, this is the same as the transformation rule for a 
genuine tensor with four contravariant indices.  However, the defining property, Eq. (\ref{defininga}), of
the Lorentz transformation only requires the square of the determinant to be $+1$, so the determinant can be $+1$ or $-1$.  If
the determinant is $-1$, then the transformation rule in Eq. (\ref{ruleforLevi}) differs from that of a genuine tensor by a change in sign. The literature of tensor analysis refers to any entity that obeys such a  rule as a pseudo-tensor, or a tensor density\cite{Bergmann}.  Here, to distinguish such from other types of 
pseudo-tensors, it is referred to as a \emph{pseudo-tensor of the standard kind}.
 
 Since $d^{\alpha\beta\mu\nu}$ is a pseudo-tensor of one kind, while $\Phi_{\mu\nu}$
is a pseudo-tensor of another kind, it follows from the basic rules of tensor calculus that their tensorial summed product in Eq.~(\ref{dual}) must be a psuedo-tensor of yet another kind where the pseudo-tensor coefficient is the product of  $\mathrm{sign}\left(\Lambda^4_4\right)$
and the sign of the determinant.   Thus   $\Psi^{\alpha\beta}$  obeys the transformation rule
  \begin{equation} \label{pseudo}
  {  \Psi}_T^{\alpha'\beta'}  =  \mathrm{sign}\left(\Lambda^4_4\right) (\mathrm{det}
[\Lambda])\Lambda^{\alpha'}_\alpha
\Lambda^{\beta'}_\beta \Psi^{\alpha\beta}.
\end{equation}

   Under  time-reversals and under all Lorentz transformations of the ortho\-chro\-nous proper sub\-group, $\Psi^{\alpha\beta}$
   transforms as  a genuine tensor, \ since the quantity
$ \mathrm{sign}\left(\Lambda^4_4\right) (\mathrm{det}
[\Lambda])$  in such instances is $+1$.  That factor is $-1$, however, for pure spatial-inversions.
 Consequently, a tensor that satisfies the rule in Eq.~(\ref{pseudo}) is here referred to as a \emph{pseudo-tensor of the spatial-inversion kind}.

Thus, the development leads to two pseudo-tensors, one which doesn't transform properly under time-reversals, and the other which
doesn't transform properly under spatial-inversions.
In principle, there is no reason why pseudo-tensors cannot be
used equally as well as genuine tensors in developing a covariant
formulation.  One must, of course, adhere to the rules of tensor calculus for categorizing tensorial products of pseudo-tensors (or
genuine tensors) of different kinds.  A sum of a pseudo-tensor and
a genuine tensor is disallowed, and so also a sum of two
 pseudo-tensors of different kinds.
   Equating any kind of pseudo-tensor  to the corresponding null tensor would be covariant, 
since a null tensor can be regarded as being whatever kind of pseudo-tensor one wishes it to be.

\section{Maxwell's equations in free space}
\label{free_space}

One now asks what determines, within the context
of a given admissible coordinate system, the time evolution
of the electromagnetic fields.  At a given point in space, the
simplest assumption is that  the momentary change in time of such fields
  depends only on the present values of those fields
in the immediate region of the point if there are no sources nearby.
Change is expected to result from imbalances, so spatial
gradients  are relevant.  
 One assumes, in the absence of
any other knowledge, that space has no intrinsic property,
other than the speed of light $c$, so one seeks a covariant
formulation introducing no further constants.
All this suggests that one seek first order partial
differential equations, which, whatever they may be,
are expressible in covariant form.
   The ordered set of
derivative operators
$\partial /\partial X^\alpha$ transforms in the same manner
as does a covariant vector, so the natural candidates for a
covariant formulation of a set of first order partial equations
are the equations:
\begin{equation} \label{firstordera}
 \frac{\partial}{\partial X^\alpha} \Phi^{\alpha\beta} = 0;
\qquad
 \frac{\partial}{\partial X^\alpha} \Psi^{\alpha\beta} = 0 .
\end{equation} 
In both cases, one can show explicitly from the
transformation rules that, if the left side is identically zero in
the reference coordinate system, then it is also zero in any
equivalent coordinate system (related to the first by any
Lorentz transformation of the full Lorentz group).  That the quantities
$\Phi^{\alpha\beta}$ and $\Psi^{\alpha\beta}$ are pseudo-tensors rather than  genuine
tensors is of no import, because the right sides can be regarded as
null tensors of the same kind.

 These, when written out explicitly, yield
 \begin{equation} \label{freeMaxa}
\nabla \times \mathbf{B} - \frac{1}{c^2}\frac{\partial \mathbf{E}}{
\partial t} = 0 ;
 \qquad  \nabla\cdot \mathbf{E} = 0;
\end{equation}  
\begin{equation}  \label{freeMaxc}
\nabla \times {\bf E}  +\frac{\partial \mathbf{
B}}{\partial t} = 0;
\qquad
 \nabla\cdot \mathbf{B} = 0,
\end{equation} 
and these are
recognized as Maxwell's equations (in rationalized MKS or
SI units) in
free space with the absence of sources.

\section{Maxwell's equations with source terms}
\label{source_terms}

The generalization of Eqs.~(\ref{firstordera})   
to allow for the presence of
localized sources is achieved by putting terms that are
possibly nonzero on the right sides, so that these become:
\begin{equation} \label{inhomogena}
\frac{\partial}{\partial X^\alpha} \Phi^{\alpha\beta} =
G^\beta ;
\qquad \frac{\partial}{\partial X^\alpha} \Psi^{\alpha\beta }=
H^\beta.  
\end{equation}
Here to achieve covariance, the right sides must transform as
the appropriate kinds of pseudo-vectors:
\begin{equation} \label{Grule}
G_T^\beta = \mathrm{sign}(\Lambda_4^4) \Lambda_\gamma^\beta G^\gamma,
\end{equation}
\begin{equation} \label{Hrule}
H_T^\beta = \mathrm{sign}(\Lambda_4^4)(\mathrm{det}
[\Lambda]) \Lambda_\gamma^\beta H^\gamma.
\end{equation}

With some hindsight, the symbols depicting the components of
the source four-vector
$G^\alpha$ are here selected to be
\begin{equation} \label{G4vector}
G^1 =  \mu_oj_x, \quad G^2 = \mu_oj_y, \quad G^3 = 
\mu_oj_z,
\quad G^4 =  \mu_o c \rho  
\end{equation}
Tentatively,
$\rho$ corresponds to charge per unit volume and the
 $j_i$ correspond to the Cartesian components of a
charge flux vector.   Note that, in regard to the transformation
rule of Eq. (\ref{Grule}),  the components of this pseudo-vector transform under time-reversals or spatial inversions as
\begin{equation}
\mathbf{j} \rightarrow -\mathbf{j},\; \qquad \rho \rightarrow \rho.
\qquad \mbox{(time-reversals or spatial-inversions)}.
\end{equation}

In contrast, were the components of the pseudo-four-vector
$H^\alpha$ to be written down in a comparable form, with
perhaps some constant different than $\mu_o$, and with
a ``magnetic charge'' flux vector $\mathbf{j}_m$ and a
``magnetic charge'' density $\rho_m$, the corresponding
transformation rules, in accord with Eq. (\ref{Hrule}), are
\begin{equation}
\mathbf{j}_m \rightarrow \mathbf{j}_m,\; \qquad \rho_m \rightarrow -\rho_m,
\qquad \mbox{(time-reversals or spatial-inversions)}.
\end{equation}
 The latter, however, presents philosophical problems.
 If magnetic charge is to be an ingredient of Maxwell's
 equations then it should be an invariant, and
 $\rho_m$ should transform into itself under time-reversals
 and spatial-inversions.   If the equations
 are not required to be covariant under time-reversals and
 spatial-inversions, then magnetic charge can be considered an invariant.  But if they are to be covariant under either, not necessarily
 both, then magnetic charge has to left out of Maxwell's equations.  [There is, however,
 a possibility that the equations can be invariant under these
 transformations, but only when they are applied simultaneously, 
 yielding a total inversion.  Such corresponds to the proper
 subgroup of Lorentz transformations, where $ \mathrm{det}
[\Lambda]=1$.]

    The formulation in the present paper began with the assumption of
a test particle that  has only one scalar property, other than
mass, this being electric charge and which is presumed to be
invariant under all Lorentz transformations.  The consideration
of sources other than those that are charge related would
therefore appear to be outside the scope of the present paper. 
Insofar as experience indicates that magnetic charge is either
nonexistent or extremely rare, or else doesn't often interact with
electric charge, and that invariance under time
reversals and spatial-inversions has intrinsic intuitive appeal, the remainder of the paper
proceeds with the assumption that $H^\alpha$ is identically
zero.   

In the symbol assignments  of Eq. (\ref{G4vector}),
 the quantity $\mu_o$ is a nonzero
constant that one is free to select.  For rationalized MKS
(SI) units, $\mu_o = 4\pi \times 10^{-7}$ $\mathrm{N/A}^2$, 
where the stated units are newtons per ampere squared
or, equivalently, henrys per meter.
[The reasons for this choice are primarily historical\cite{Weil}${}^,$\cite{Bailey}${}^,$\cite{Taylor}.  The
resulting Maxwell's equations, in either cgs or MKS units, in
conjunction with the continuum extension of the Lorentz
force law, yield the magnetostatic result
\begin{equation} \label{magnetostat}
d{\bf F}_{12} = \frac{\mu_o}{4\pi} I_1 I_2
\left( d\mathbf{ l} \times \int \frac{ d\mathbf{l}^\prime \times
(\mathbf{r} - \mathbf{r}^\prime)}{|\mathbf{r} - \mathbf{r}^\prime|^3 }
\right) .
\end{equation}
Here $d\mathbf{ F}_{12}$ is the incremental force exerted by the
electrical current $I_1$ in a thin wire on a length element 
$d\mathbf{l}$ of a second wire that is carrying a current $I_2$.
The line integral passes along the closed circuit of the first wire
in the direction of current flow.  The element $d\mathbf{l}$ is at
the point $\mathbf{r}$, and $\mathbf{r}^\prime$ denotes points on 
the first wire.  In the original system of electromagnetic
units, the unit of current, subsequently termed the abampere,
was defined so that the coefficient  $\mu_o/4\pi$ in the
above formula was unity.  Since force was in dynes, this
rendered $\mu_o$ equal to $4\pi$ dynes per abampere
squared.  An international agreement in 1881 fixed the
magnitude of the ampere to be $0.1$ abampere.  Since the
dyne is $10^{-5}$ newtons, one has 1 dyne per abampere
squared equal to $10^{-7}$ newtons per ampere squared, and
hence the numerical value $4\pi\times 10^{-7}$ results.  This
assignment  and Eq.~(\ref{magnetostat}) yield 
the standard definition of the ampere,
coulombs per second, in terms of the hypothetical experiment:
two parallel straight wires each carrying a current $I$ are
placed one meter apart.  If $I$ is one ampere, then the force
per unit length exerted on one wire by the other is $2\times
10^{-7}$ newtons per meter.]

With the additional
introduction of a symbol $\epsilon_o$, defined as 
\begin{equation} \label{epsnaught}
 \epsilon_o =\frac{ 1}{\mu_o c^2},
 \end{equation}
 the Eqs.~(\ref{inhomogena}),
when written out explicitly and expressed in vector notation,
yield:
\begin{equation}  \label{Maxwella}
 \nabla \times (\mathbf{B}/\mu_o) - \frac{\partial (\epsilon_o\mathbf{
E})}{ \partial t} = {\bf j};
\end{equation}
\begin{equation} \label{Maxwellb}
\nabla\cdot(\epsilon_o \mathbf{E}) =  \rho; 
\end{equation}
\begin{equation} \label{Maxwellc}
 \nabla \times \mathbf{E} +\frac{\partial \mathbf{B}}{\partial t} =
0;
\end{equation}
\begin{equation} \label{Maxwelld}
\nabla\cdot \mathbf{B} =  0. 
\end{equation}
These equations,
within which the speed of light $c$ does not explicitly appear,
have the form of Maxwell's equations  in rationalized MKS
(SI) units that one commonly sees\cite{Stratton} in the literature, only with
the substitutions in the first two of these of
\begin{equation}
 \mathbf{B}/\mu_o = {\bf H}; \qquad \epsilon_o \mathbf{E} = \mathbf{D}.
\end{equation}

The partial differential equation for the conservation of charge,
 \begin{equation}  \label{currentcons}
 \nabla\cdot \mathbf{j} + \frac{\partial \rho}{\partial t} = 0 ,
\end{equation}
 follows from Eqs.~(\ref{Maxwella}) and (\ref{Maxwellb}), because the
divergence of the curl of ${\bf B}$ is zero. An alternate
derivation recognizes that, because $ \Phi^{\alpha\beta}$ is
antisymmetric, one has
\begin{equation}
 \frac{\partial}{\partial X^\beta} \frac{\partial}{\partial
X^\alpha } \Phi^{\alpha\beta}
 =0  , 
\end{equation} 
which, in conjunction with Eq.~(\ref{inhomogena}),
yields
\begin{equation}
\frac{\partial G^\beta }{ \partial X^\beta} = 0,
\end{equation}
and this, with the symbol assignments in Eq.~(\ref{G4vector}),
yields Eq.~(\ref{currentcons}).

\section{Equivalence of types of charge}
\label{types_of_charge}

 The association of the quantities $\rho$ and $\mathbf{j}$ in the
 source terms with charge per unit volume and with flux
 of moving charge can, at one level, be regarded as a postulate
 and as merely defining the units of charge.   Such, however,
 may have little intuitive appeal, and it is consequently desirable
 to appeal to some principle that seems intrinsically more
 plausible.   To this purpose, reference is made to Newton's
 original enunciation\cite{Newton}  of his third law:
 
 \begin{description}

\item{ \textbf{Law 3.} }{\emph{ To every action there is always an opposite
and equal reaction; in other words the actions of two bodies upon each
other are always equal and always opposite in direction.}}

 \end{description}
 
 To apply this law to the present circumstances, it is necessary to interpret the
 word \emph{always} as meaning ``in all instances'' rather than 
 ``at every moment of time.''  Insofar as forces are transmitted instantaneously
 from one body to another or else are constant in time, the usual interpretation 
  applies with action
 interpreted as the vector force.  If the finite time of propagation of changes
 in force is to be taken into account, then the suggested replacement is
 that time integrals of forces be equal and opposite.     This is consistent
 with the general idea of what is often stated as the \emph{principle
 of reciprocity};  effect per unit source strength of a source on a small
 ``effect receiver''  is  the same when the roles, ``source'' and
 ``effect receiver'' 
 are interchanged\cite{Landauelectro}${}^,$\cite{Morsevolone}.  Here the terminology is intentionally
 vague;  a precise statement, based on the hint provided by
 Feynman\cite{Dyson}, is
 
 \begin{quote}
 
 \emph{The equations that govern the electromagnetic interaction  
 between electrical charges must be such that, if a given moving
charge is the source of an electromagnetic field that exerts a force on a second moving charge, then the same equations (Maxwell's equations with source terms plus the Lorentz force law)  apply for the
determination of the force  exerted on the first charge by the electromagnetic field caused by the second charge.}
 
 \end{quote}
 
 \noindent Implicit in this is that one can define units for charge so that 
 all charges, whether sources or recipients of force, should have the same units.
 
 To demonstrate that this reciprocity statement leads to the
 requirement of equal and
 oppositely directed time-integrals of forces,
  one begins with the hypothesis of $\rho$ and
 $\mathbf{j}$ being appropriately interpreted in the manner stated above.
    Because one is
 here concerned with point particles, one expresses $\rho$ 
 and $\mathbf{j}$ as sums over point particles,
 \begin{equation}\label{chargesums}
 \rho = \sum_Q q_Q \delta(\mathbf{x} 
- \mathbf{x}_Q); \qquad
 \mathbf{j} = \sum_Q \mathbf{v}_Q q_Q \delta(\mathbf{x} 
- \mathbf{x}_Q)
\end{equation}
Here $q_Q$ is the magnitude (possibly negative) of the $Q$-th
charge,  $\mathbf{x}_Q$ is its position vector, and 
  $\mathbf{v}_Q$ is its velocity vector. The quantity
$\delta(\mathbf{x}  - \mathbf{x}_Q)$ is the three-dimensional
$\delta$-function, defined so that it is singularly concentrated at
the instantaneous location of the particle and so that its volume
integral is unity.  The conservation of charge equation,
Eq.~(\ref{currentcons}), holds trivially with these identifications, 
since
\begin{equation}
\left(\mathbf{v_Q}\cdot\nabla +\frac{\partial}{\partial t}\right)\left(\mathbf{x} 
- \mathbf{x}_Q\right) = 0.
\end{equation}
Also, the four-vector $G^\alpha$, with the substitution
 of Eqs.~(\ref{chargesums})
into Eq.~(\ref{G4vector}),  continues to transform as a pseudo-four-vector of the time-reversal type.
To verify this, one first notes that the substitution renders
 \begin{equation}\label{sumcharges}
G^\alpha =\mu_o \sum_Q q_Q U_Q^\alpha(\tau_Q) \delta(\mathbf{x} 
- \mathbf{x}_Q)\frac{d\tau_Q}{dt}.
\end{equation}
 where $U_Q^\alpha$ is the $Q$-th charge's four-vector velocity vector,
 and $\tau_Q$ is its proper time. One  then
 notes that the quantity $ \delta(\mathbf{x} 
- \mathbf{x}_Q){d\tau_P}/{dt}$ transforms as a scalar, because
${d\tau_P}/{dt}$ is invariant under time-reversals, because
the Jacobian for changing from one Minskowki space to another
in a four-dimentional integration is unity (recall that 
the determinant of the Lorentz transformation is always 1 or -1),
and because the time integration over a transformed time integral
of ${d\tau_Q}/{dt}$ is $\Delta \tau_Q$, which is an invariant.
The quantity $U_Q^\alpha$ is a pseudo-four-vector of the time-reversal kind, and a scalar times
such a four-vector is also a pseudo-four-vector of the same kind. 

Then, to derive a relation with a resemblance 
to Newton's third law,
one conceives of  fields, $\mathbf{E}_Q$ and $\mathbf{B}_Q$,
caused by charge $q_Q$, these satisfying the equations
\begin{equation}  \label{MaxwellQa}
 \nabla \times (\mathbf{B}_Q/\mu_o) - \frac{\partial (\epsilon_o\mathbf{
E}_Q)}{ \partial t} = q_Q{\bf v}_Q\delta(\mathbf{x} 
- \mathbf{x}_Q);
\end{equation}
\begin{equation} \label{MaxwellQb}
\nabla\cdot(\epsilon_o \mathbf{E}_Q) =  q_Q \delta(\mathbf{x} 
- \mathbf{x}_Q); 
\end{equation}
\begin{equation} \label{MaxwellQc}
 \nabla \times \mathbf{E}_Q +\frac{\partial \mathbf{B}_Q}{\partial t} =
0;
\end{equation}
\begin{equation} \label{MaxwellQd}
\nabla\cdot \mathbf{B}_Q =  0. 
\end{equation}
The force exerted on charge $q_P$ because of the influence of 
charge $q_Q$ is consequently
\begin{equation}
\mathbf{f}_{PQ} =  q_P\mathbf{E}_Q + q_P \mathbf{v}_P \times
\mathbf{B}_Q,
\end{equation}
where the two fields are understood to be evaluated at the
position, $\mathbf{x}_P$, of the charge $q_P$.
If charge $q_P$ should also be the source of a field that exerts
an influence on charge $q_Q$, then there must be analogous 
relations that result from the above with the interchange of the
subscripts $P$ and $Q$.

A derivation, analogous to what one finds often used for proving the
invariance of Green's functions\cite{Morsevolone} under reciprocity,
 proceeds  from the Maxwell equations for the fields $\mathbf{B}_Q$,
 $\mathbf{E}_Q$, $\mathbf{B}_P$, and $\mathbf{E}_P$,  and yields
 the result
 \begin{eqnarray} \label{reciprocity}
\lefteqn{ \sum_{ij}\frac{\partial M_{PQ,ij}}{\partial x_i} \mathbf{e}_j
 + \frac{\partial \mathbf{N}_{PQ}}{\partial t}  }
 \nonumber\\
   & = q_P\left(\mathbf{E}_Q + \mathbf{v}_P\times\mathbf{B}_Q\right)
 \delta(\mathbf{x} - \mathbf{x}_P)
  +q_Q\left(\mathbf{E}_P + \mathbf{v}_Q\times\mathbf{B}_P\right)
 \delta(\mathbf{x} - \mathbf{x}_Q).
 \end{eqnarray}
 [This is a special case of a more general relation previously derived by Goedecke.\cite{Goedecke}]
 The most important feature of the left side of this equation, from
 the standpoint of the present discussion, is that it is a sum of
 derivatives.  The abbreviated differentiated quantities are
 \begin{eqnarray}
\lefteqn{ M_{PQ,ij} = \frac{1}{\mu_o} \left(-\sum_k B_{P,k}B_{Q,k}\delta_{ij}
 + B_{P,j}B_{Q,i} + B_{P,i}B_{Q,j}\right) }
 \nonumber\\
& +\epsilon_o \left(-\displaystyle{\sum_k} E_{P,k}E_{Q,k}\delta_{ij}
 + E_{P,j}E_{Q,i} + E_{P,i}E_{Q,j}\right) ,
 \end{eqnarray}
 \begin{equation}
 \mathbf{N}_{PQ} = \epsilon_o\left( \mathbf{B}_P \times \mathbf{E}_Q
 + \mathbf{B}_Q \times \mathbf{E}_P\right).
 \end{equation}
 
 Integrating both sides of Eq.~(\ref{reciprocity}) over the volume of a large sphere
 surrounding the two charges, using Gauss's theorem to convert some
 volume integrals to surface integrals, and then letting the radius of the sphere
 approach infinity, yields the result
 \begin{equation}\label{integratedrecip}
 \mathbf{f}_{PQ} + \mathbf{f}_{QP} = \frac{d}{dt}\int_V  \mathbf{N}_{PQ} dV.
 \end{equation}
 [The vanishing of the surface integrations at infinite radius is not trivially true in all instances, but its plausibility is evident when one considers the transient
 case and allows that the disturbances propagate at a finite speed, so
 that they never reach infinity.]
 
 Equation (\ref{integratedrecip}) is the precise covariant statement
 of Newton's third law in the present context, and its emergence justifies
 the physical equivalence of the charge in source terms to that of the test particle.
 If the particles are stationary, then the forces are equal and opposite,
 $\mathbf{f}_{PQ} =- \mathbf{f}_{QP} $.  It is beyond the scope of the
 present article to discuss all the circumstances when this is still
 a good approximation, but it is evident that time integrals over 
 extended time intervals will tend to smooth out 
 fluctuations, so that
 \begin{equation}
 \int \mathbf{f}_{PQ} dt \approx - \int  \mathbf{f}_{QP} dt.
 \end{equation}
 
 [A comparable discussion to what appears above can be found in the 
 1966 paper by Tessman\cite{Tessman}, who derived the electric and magnetic fields of an accelerating charge, using a number of plausible assumptions, but not explicitly invoking Maxwell's equations.  One of the assumptions, replacing Newton's third law, was that ``the total electric force which a stationary charge exerts upon a system of charges in steady state is equal in magnitude and opposite in direction to the total electric force exerted by the steady state system 
 upon the steady charge.''  This apparently was sufficient, although
 from the standpoint of the special theory of relativity, it is
 desirable to have a statement that does not require the existence
 of a special coordinate system in which all the relevant charges are
 stationary.  Another assumption which Tessman makes is that the force exerted on a second charge at a given time is due only to
 the first charge's dynamical state at the retarded time $t-(r/c)$,
 where $r$ is the distance from the first charge's position at that
 time to the second charge's position at the current time.  Such is
 more in keeping with the stronger use of the principle of reciprocity,
 but the statement can become unwieldy for general formulations when one considers that there may be more than one point on the
 first charge's trajectory that meets this criterion.]

There is still one further requirement to be satisfied:  both $q_P$ and $q_Q$ must have the same units. 
This is guaranteed if the numerical values that are assigned to
both charges are measured in a consistent manner. 
Consideration of the case of only two charges cannot resolve
this, as the forces depend only on the product of the two
charges.  One can calibrate the charges, however, if one has a
third charge $q_3$.  The value of $q_3$ need not be known, but
the forces exerted by it on $q_P$ and $q_Q$  in the static
limit, in conjunction
with Coulomb's law  (which is derivable from the equations given above), 
enable one to determine the ratio of the
two charges.  Then the Coulomb's law relation for the force between
the two suffices to determine the magnitude of either
$q_P$ or $q_Q$.  Given the choice of $\epsilon_o$ represented
by Eq.~(\ref{epsnaught}), this would 
determine the numerical value
in coulombs of either charge.  One does not necessarily
measure charge in this manner, but the mere fact that some
measurement procedure exists insures that one can always
take the two charges to have the same units.

\section{Concluding remarks}
\label{Concluding}

 The present paper presents an alternate derivation of a 
  standard result, i.e., Maxwell's equations. Whether it provides significant new insight, a significantly
  new way of thinking, or a much simpler approach, is,
  as Krefetz\cite{Krefetz}  wrote many years ago in a similar context,
  ``after all, a matter of taste.''    Newton's laws are an attractive 
  starting point, as they seem the most intuitively appealing of
  all the laws of physics, even though the idea of their universal applicability has long since been abandoned.  Their vestiges
  remain in practically all of current physics, and it is difficult to conceive of  a curriculum in physics or in one of the many branches of applied physics that does not begin with Newton's laws.  
  
Various options were left open throughout the derivation as to
what is meant by the ``covariance requirement.''  In the bulk
of the literature on special relativity, covariance is implicitly
understood to mean covariance under the orthochronous 
proper Lorentz group, and thereby such literature has 
implicitly ignored the
possibility of achieving or not achieving covariance under
time reversals and spatial inversions.  As Dixon\cite{Dixon}
points out: ``[Although] attention is normally restricted to
coordinate systems in which the time coordinate increases
into the future and in which the spatial coordinates are right
handed, this is very different from the question of whether the
laws of physics themselves determine a particular orientation
or time-orientation in spacetime. \ldots However, no fundamental
law outside the domain of quantum physics has yet shown such
an asymmetry.\ldots  [Although] it is inconvenient to develop the
laws of physics without ever making definitions which depend
on an arbitrary orientation or time-orientation, \ldots the
behavior of equations under a change of convention is
important.''   In this spirit, the development in the present paper
has been careful to specify the manner in which quantities such
as the four-vector particle velocity and the two electromagnetic
field tensors transform under the full Lorentz group.   

One difference between the treatment here from that in many treatments of electromagnetism is that no potentials are introduced.
In retrospect, the setting of the four-vector source term $H^\alpha$
to zero in the differential equations, Eq. (\ref{inhomogena}), is equivalent
to assuming that such potentials exist.   [See, for example,
Exercise 9 on page 140 of the text by Synge and Child.\cite{Synge}]
The relevant question is what is intrinsically more plausible. The
development here rests on a presumed symmetry in spacetime.
If the equations are to be covariant under time-reversal or if they
are to be covariant under spatial inversion, then potentials exist.
In some literature, the competing argument is that magnetic
monopoles do not exist.  The argument here is that magnetic charge is precluded in source terms in equations of macroscopic physics if
such equations are to be covariant under either spatial inversions or
time-reversals (not necessarily both).  In various places in the literature, one finds one or the other mentioned as precluding magnetic charge.  In retrospect, it is clear why either type of covariance suffices, as $\nabla\cdot\mathbf{B}$ changes sign
under either time-reversal or spatial inversion.

In this respect, one may note an intriguing remark made some
time ago by Schiff\cite{Gold}: ``It is usually said that Newton's
laws and Maxwell's equations are time-reversible.  These are
time-reversible if there are no charges but no monopoles or
if there are monopoles but no charges, but not if there are both.''
The context does not make it clear exactly what Schiff meant, but
the development here yields an interesting interpretation.  The
analysis here started with the hypothesis of the existence of a
test particle with a scalar property.  The ensuing result was that
the fields that affect such a test particle are caused by the presence
of other particles with the same type of scalar property.  Perhaps
other types of fields exist, such as are caused by some particles
with a different type of scalar property.  Given full covariance, our test particle cannot sense their presence.  Suppose, on the other hand, one started with a test particle that had a scalar property which one
chose to term ''magnetic charge.''  The same equations will result
but one can always choose the symbols for the elements of the
field tensors so that the new Lorentz force will involve a 
linear combination\cite{Rindler} of terms such as $q_m \textbf{B}$ and $q_m\textbf{v}_P\times\textbf{E}$.  But in this case, time reversal
covariance or spatial inversion covariance will preclude the
presence of source terms which involve electric charge.
Thus, something like magnetic charge could very well exist, but at
the macroscopic level its presence cannot be sensed in terms of
forces on particles with electric charges, given that the equations
are to be fully covariant.


\begin{thebibliography}{99}

 
 
 
 \bibitem{Einstein1905}  Einstein, A.:
 Zur Elektrodynamik bewegter K\"orper.  Ann.  Physik
 17, 891--921 (1905)
 
 \bibitem{DoverRelativity} Lorentz, H. A.,  Einstein, A., Minkowski, H.,  Weyl, H.:  The Principle
of Relativity.   Dover, New York  (1952) 

\bibitem{Kilmister}  Kilmister, C. W.:  Special Theory of Relativity.  Pergamon, Oxford  (1970) 


  
\bibitem{Minkowski1908} Minkowski, H.:   Die Grundgleichungen f\"ur
die electromagnetischen Vorg\"ange in bewegten K\"orpen.
Nachtrichten der K. Gesellschaft der Wissenschaften zu 
G\"ottingen.  Mathematisch-physikalische Klasse: 53--116 (1908).
 





 \bibitem{Dyson} Dyson,  F. J.: Feynman's proof of the Maxwell
equations.  Am. J. Phys. 58(3), 209--211 (1990)

\bibitem{Page1912}  Page, L.:  A derivation of the fundamental relations
of electrodynamics from those of electrostatics.  Am. 
J.  Sci.  34(199), 57--68 (1912)




\bibitem{PageandAdams} Page, L.,  Adams, N. I., Jr.  Electrodynamics, pp. 129--154.  
Van Nostrand, New York  (1940) 

\bibitem{Frisch} Frisch, D. H., Wilets, L.:  Development of the
Maxwell-Lorentz equations from special relativity and Gauss's
law.   Am. J. Phys. 24, 574--579  (1956)




\bibitem{Swann} Swann, W. F. G.:  New deductions of the electromagnetic equations.  Phys. Rev. 28(3),
531--544 (1926)

\bibitem{ElliottIEEE} Elliott, R. S.:  Relativity and electricity. IEEE
Spectrum  3, 140--152 (March 1966)


\bibitem{Tessman} Tessman, J. R.:  Maxwell --- Out of Newton,
Coulomb, and Einstein.  Am. J. Phys. 34, 1048--1055 (1966)






\bibitem{Krefetz} Krefetz, E.:   A ``derivation'' of Maxwell's equations.
 Am. J. Phys.   38(4), 513--516  (1970)
 
 \bibitem{Feynmanbook} Feynman, R. P.:
 Lorentz transformations of the fields.  In:
  Feynman, R. P.,  Leighton, R. B.,  Sands, M. (eds.)The Feynman Lectures on Physics, mainly
 Electromagnetism and Matter, p. 26-2.   Addison-Wesley, Reading (1964) 
  
 \bibitem{Podolsky}  Podolsky, B.,  Kunz, K. S.: Fundamentals of
Electrodynamics, pp. ix,
101--124.
 Marcel Dekker, New York  (1969) 
 
\bibitem{LandauFields}  Landau, L. D., Lifshitz, E. M.:   The Classical
Theory of  Fields, pp. 14-19, 44--46, 60--62, 66--75.
Pergamon, Oxford  (1975) 

 \bibitem{Kobe1978}  Kobe, D. H.:   Derivation of Maxwell's equations from
the local gauge invariance of quantum mechanics.  Am. J. Phys.
46(4), 342--348  (1978)



\bibitem{Kobe1980}   Kobe,  D. H.:    Derivation of Maxwell's equations from
the gauge invariance of classical mechanics.   Am. J. Phys. 
48(5), 348--353  (1980)



\bibitem{Kobe1984}  Kobe, D. H.: ``Helmholtz theorem for
antisymmetric second-rank tensor fields and electromagnetism with
magnetic monopoles.  Am. J. Phys.  52(4), 354--358 (1984)

\bibitem{Kobe1986}   Kobe, D. H.:   Generalization of Coulomb's law to
Maxwell's equations using special relativity.  Am. J. Phys. 
54(7), 631--636 (1986)

\bibitem{Crater}  Crater,   H. W.:  General covariance, Lorentz
covariance, the Lorentz force, and Maxwell's equations.  Am. J.
Phys.  62(10), 923--931 (1994)

\bibitem{Jefimenko1996}   Jefimenko,  O. D.:  Derivation of relativistic force
transformation equations from Lorentz force law.  Am. J.
Phys. 64(5), 618--620 (1996)

\bibitem{Ton}  Ton, T.-C.:   On the time-dependent, generalized
Coulomb, and Biot-Savart Laws,  Am. J. Phys.   
59(6), 520--528
(1991)

\bibitem{Griffiths}  Griffiths, D. J., Heald,  M. A.: Time-dependent
generalizations of the Biot-Savart and Coulomb laws.   Am. J.
Phys.  59(2), 111-117 (1991)

\bibitem{Crawford}  Crawford,  F. S.:   Magnetic monopoles, Galilean
invariance, and Maxwell's equations.   Am. J. Phys.  60(2),
109--114  (1992)

\bibitem{Neuenschwander}  Neuenschwander, D. E., Turner, B. N.:
Generalization of the Biot-Savart law to Maxwell's
equations.   Am. J. Phys.   60(1), 35--38  (1992)

\bibitem{Bork} Bork, A. M.:  Maxwell, displacement current, and symmetry.
Am. J. Phys.  31, 854--859 (1963)

\bibitem{Goedecke} Goedecke, G. H.:   On electromagnetic conservation laws.   Am. J. Phys.  68(4), 380--384 (2000)

\bibitem{Hokkyo}   Hokkyo, N.:   Feynman's proof of Maxwell equations and Yang's unification of electromagnetic and gravitational Aharonov-Bohm effects.   Am. J. Phys.   72(3),
345--347 (2004)

 


\bibitem{Newton}  Newton, I.:   The Principia:  Mathematical Principles
of Natural Philosophy,  pp. 416--417.
Univ. Cal. Pr., Berkeley  (1999) 


\bibitem{Minkowski1909} Minkowski, H.:``Raum und Zeit,''
Phys. Zeit. 10, 104--111 (1909)   


\bibitem{Bergmann}  Bergmann, P. G.:  Introduction to the Theory of
Relativity, pp. 47--120.  Dover, New York (1976) 

\bibitem{EinsteinGrundlage1916}   Einstein, A.: Die Grundlagen der allgemeinen
Relativit\"atstheorie.  Ann.  Physik, ser. 4, 49(7),
769--822 (1916)

\bibitem{Low} Low, F. E.:  Classical Field Theory:  Electromagnetism
and Gravitation,  pp. 252--255, 259--260, 269--270.  Wiley, New York (1997)



\bibitem{Cunningham}   Cunningham, E.:
The principle of relativity in electrodynamics and an extension
thereof.    Proc. London Math. Soc.  8, 77--98 (1909).

\bibitem{Bateman} Bateman, H.:
The transformation of the electrodynamical equations.
Proc. London Math. Soc.   8(2), 223--264 (1910)



\bibitem{Poincare1906}   Poincar\'e, H.: Sur la dynamique de l'\'electron.
Rendiconti del Circolo matematico di Palermo 21,
129--176 (1906)
 
\bibitem{SchwartzRendiconti} Schwartz, H. M.: Poincar\'e's
Rendiconti paper on relativity:  Part I.  Am. J. Phys. 39(11),
1277--1294 (November 1971).  Part II, ibid. 40(6),
862--872 (June 1972).   Part III, ibid., 40(9), 1282--1287
(September 1972).

\bibitem{Weinberg}  Weinberg, S.:  The Quantum Theory of
Fields, vol. 1, pp. 55--58.  Cambridge Univ. Pr. (1995) 

\bibitem{Schwartz} Schwartz, M.:  Principles of Electrodynamics, pp. 127--129.
Dover, New York (1987) 

\bibitem{Heaviside} Heaviside, O.:  On the electromagnetic effects due to the motion of electrification through a dielectric.
Phil. Mag.  27, 324--339 (1889) 

\bibitem{Lorentzbook}  Lorentz, H. A.: The Theory of Electrons, p. 14.
Dover, New York (1952) 

\bibitem{Tolman}  Tolman,  R. C.:   Note on the derivation from the
principle of relativity of the fifth fundamental equation of the
Maxwell-Lorentz theory.   Phil. Mag., ser. 6, 21(123), 296-301 (March 1911)





\bibitem{Schilpp}   Einstein, A.:  Autobiographical notes.  In: Schilpp, P.  A. (ed.)
 Albert Einstein:
Philosopher-Scientist, p. 63.  Tudor Publ., New York (1951)

 \bibitem{Moller}  M\o ller, C.:  The Theory of Relativity, pp. 113--114.   Oxford Univ. Pr.
(1952) 

\bibitem{Levi}   Levi-Civita, T.:The Absolute Differential
Calculus, pp. 158--160.  Dover, New York  (1977) 

\bibitem{Synge}  Synge, J. L., Schild, A.:  Tensor Calculus, pp. 131--135. 
 Dover, New York  (1978) 

\bibitem{Schreier} O. Schreier, O.,  Sperner, E.:
Modern Algebra and Matrix Theory, p. 89. Chelsea,
New York (1955)



 


\bibitem{Stratton}  Stratton, J. A.: Electromagnetic Theory, pp. 2--6.  McGraw-Hill, New York (1941) 

\bibitem{Landauelectro} Landau, L. D., Lifshitz, E. M.:  Electrodynamics of Continuous Media, pp. 288-289.   Pergamon, London (1960) 

\bibitem{Morsevolone}   Morse, P.  M. , Feshbach, H.:
Methods of Theoretical Physics,  vol. I,   pp. 804--806, 834--837.  McGraw-Hill,
New York  (1953)

\bibitem{Weil}   Weil, J. F.:   Units of measurement.  
 In:  McGraw-Hill Encyclopedia of Science and Technology, 9th
edition, vol. 19, pp. 64--72.  McGraw Hill, New York (2002) 

\bibitem{Bailey}  Bailey,  A. E.:   Electrical units and standards.   In:
McGraw-Hill Encyclopedia of Science and Technology, 9th
edition, vol. 6,  pp. 226--231.
McGraw Hill, New York  (2002) 


\bibitem{Taylor} Taylor, B. N. (ed.):  The International
System of Units (SI), pp. 1--2, 6--7, 11, 18, 32--33.
 Natl. Inst. Stand. Technol. Spec. Publ. 330.
 U. S. Government Printing Office, Washington (2001)
 

\bibitem{Dixon} Dixon, W. G.: Special Relativity:
The Foundation of Macroscopic Physics, p. 89.  Cambridge
Univ. Pr. (1978) 

\bibitem{Gold}  Gold, T.:  The Nature of Time, p. 216.
Cornell Univ. Pr. (1967) 

\bibitem{Rindler}  Rindler,  W.:  Relativity and electromagnetism: The
force on a magnetic monopole.   Am. J. Phys. 
{57}(11), 993--994
(1989)


\end{thebibliography}
\end{document}